\documentclass{article}
\usepackage{spadre2008}
\usepackage{graphicx}
\frompage{000} \topage{000}                                              
\def\rightmark{(Short) Prototype Performance of MTD at STAR}
\def\leftmark{L. Ruan et al.}

\title{Prototype Performance of Novel Muon Telescope
Detector \\ at STAR}
\authors{
{Lijuan Ruan (for the STAR Collaboration)$^1$ %
}\\[2.812mm]
{\normalsize \hspace*{-8pt}$^1$ Physics Department, Brookhaven
National Laboratory, Upton, NY 11973, USA\\[0.2ex]
}}

\abstract{Research on a large-area, cost-effective Muon Telescope
Detector (MTD) has been carried out for RHIC and for next
generation detectors at future QCD Lab. We utilize
state-of-the-art multi-gap resistive plate chambers with large
modules and long readout strips in detector design~\cite{MTDLDRD}.
The results from cosmic ray and beam test will be presented to
address intrinsic timing and spatial resolution for a Long-MRPC.
The prototype performance of a novel muon telescope detector at
STAR will be reported, including muon identification capability,
timing and spatial resolution.}

\keyword{Muon Telescope Detector, long-MRPC, intrinsic timing and
spatial resolution, muon identification capability}
\PACS{25.75.Cj, 29.40.Cs}

\begin{document}

\maketitle
\setcounter{page}{1}

\section{Introduction}\label{intro}

A large-area muon detector at mid-rapidity for RHIC collisions
will be crucial for advancing our knowledge of Quark-Gluon Plasma
(QGP) properties. It directly addresses many of the open questions
and long-term goals proposed in STAR white
papers~\cite{starwhitepaper}. Since muons do not participate in
strong interactions, they provide penetrating probes for the
strongly-interacting QGP. A compact detector identifying muons
with momentum of a few GeV/$c$ at mid-rapidity, allows for the
detection of di-muon pairs from QGP thermal radiation, quarkonia,
light vector mesons, possible correlations of quarks and gluons as
resonances in QGP, Drell-Yan production, as well as the
measurement of heavy flavor hadrons through their semi-leptonic
decays into single muons~\cite{dilepton}. Some of these topics can
also be studied using electrons or photons or a combination of
both. However, they have large backgrounds from hadron decays at
the interaction, $\pi^{0}$ and $\eta$ Dalitz decay and gamma
conversions in the detector material. These backgrounds prevent an
effective trigger in central nucleus-nucleus collisions in a
detector with large coverage. In addition to an effective trigger
and cleaner signal-to-background ratio, electron-muon correlation
can be used to distinguish lepton pair production and heavy quark
decays ($c+\bar{c}\rightarrow e+\mu(e)$, $B \rightarrow e(\mu)+ c
\rightarrow e+\mu(e)$). Besides, muons are less affected than
electrons by radiative losses in the detector materials, thus
providing excellent mass resolution of vector mesons. For example,
different Upsilon states ($\Upsilon(1S)$, $\Upsilon(2S)$ and
$\Upsilon(3S)$) can be separated through dimuon decay channel in
the invariant mass distribution.

Conventional muon detectors rely heavily on tracking stations
while this new detector proposes to use good timing ( $\!<\!100$
ps) and coarse spatial ( $\sim\!1$ cm) resolution to identify
muons with momentum of a few GeV/$c$~\cite{MTDLDRD}. The multi-gap
resistive plate chamber technology with large modules, long strips
and double-ended readout (Long-MRPC) was used for this research.
Similar technology but with small pads is being built for STAR as
a Time-of-Flight Detector~\cite{upVPD}.

\section{Simulation}\label{simu}

\begin{figure}
\vspace{3.5cm} \hspace{2.5cm}
\includegraphics[keepaspectratio,scale=0.8]{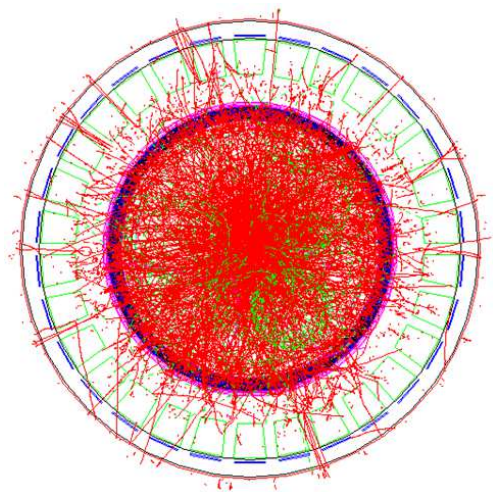}
\caption{(in color online) A full HIJING central Au+Au collisions
simulated in STAR.} \label{figure1}
\end{figure}

The simulation of a full HIJING central Au+Au collisions is shown
in Fig.~\ref{figure1} using STAR year 2003 geometry with full
configuration of the detectors and a complete material budget. We
created a muon-detector (MTD) (in blue) covering the full magnet
steel within $|\eta|\!<\!0.8$ and left the gaps in-between
uncovered, which corresponds to 56.6\% of $2\pi$ in azimuth.
Fig.~\ref{figure1} shows that most of the particles are stopped
before passing the Barrel Electromagnetic Carolimeter and most of
the escaping particles (primary or secondary) come through the
gaps in the magnet steel (in green). Further simulation with STAR
geometry indicates that for a muon track at $p_{T}\!>\!2$ GeV/$c$
generated in the center of the Time Projection Chamber (TPC), the
detection efficiency of the MTD including acceptance effect is
about 40-50\% while for a pion track, the efficiency is 0.5-1\%. A
matched MTD hit, a precise time of flight measurement from the MTD
and the current ionization energy loss (dE/dx) identification
capability from the TPC, will give us a muon-to-hadron enhancement
factor of 100-1000. Also, requiring two MTD hits in the trigger
will enhance the di-muon spectra by a factor of 10-50. This
together with data acquisition at $\!>\!1000$ Hz will greatly
enhance the capability of $J/\Psi$ and other dilepton programs in
RHIC II and future QCD Lab~\cite{MTDLDRD,DNP06}.

\section{Intrinsic timing and spatial resolution of
long-MRPC}\label{cosmicbeamtest}

\begin{figure}
\vspace{4.5cm} \hspace{2.5cm}
\includegraphics[keepaspectratio,scale=0.75]{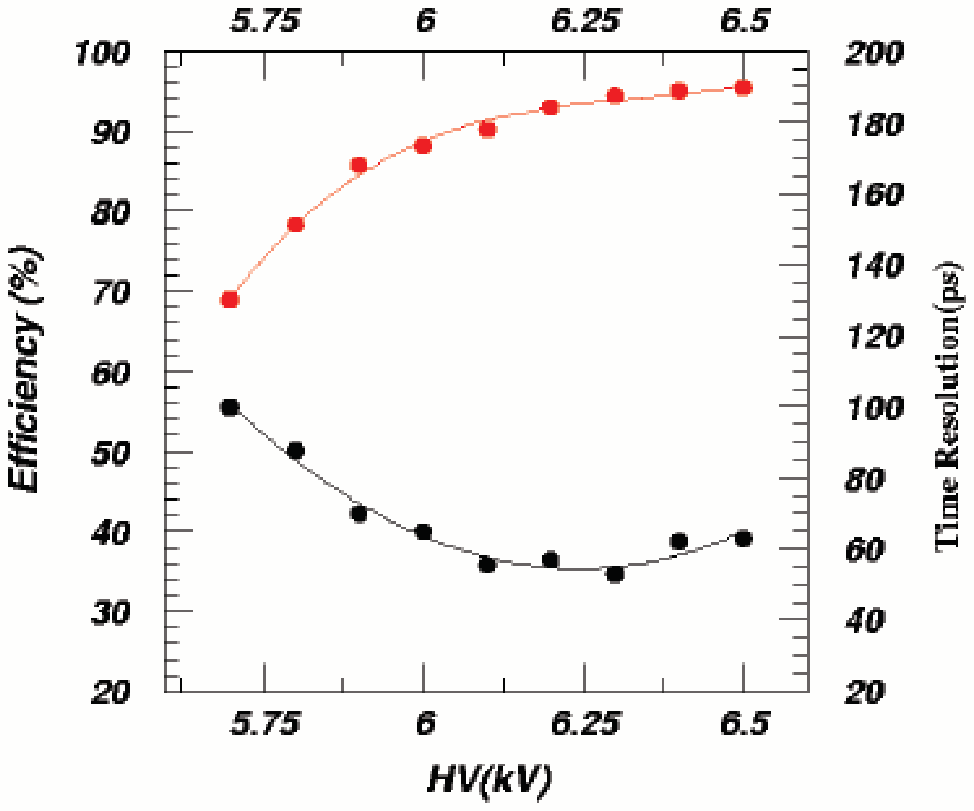}
\caption{(in color online) Efficiency and intrinsic timing
resolution, from the cosmic ray tests, as a function of half the
applied high voltage.} \label{figure2}
\end{figure}

Each long-MRPC module consists of two stacks of resistive glass
plates with ten uniform gas gaps with gap widths of 250 $\mu$m.
High voltage is applied to electrodes on the outer surfaces of the
outer plates of each stack. A charged particle traversing a module
generates avalanches in the gas gaps which are read out by six
copper pickup strips with strip dimensions of $870\times25$
$\mathrm{mm}^{2}$. The MRPC modules were operated at 12.6 kV with
a mixture of 95\% $C_{2}H_{2}F_{4}$ and 5\% iso-butane at 1
atmosphere. Fig.~\ref{figure2} shows the efficiency and intrinsic
timing resolution as a function of half of applied high voltage
(HV) from the cosmic ray test. In the high voltage range
$\!12.5\!<\!HV\!<\!13.0$ kV, the efficiency is above 95\% and
timing resolution is about 60-70 ps. In additional to the cosmic
ray test, a beam test named T963 was carried out in the MTEST beam
line at Fermi National Accelerator Laboratory (FNAL) in May 2007.
The results from beam tests using prototype front-end electronics
show timing resolution and efficiency consistent with those from
the cosmic ray tests. The spatial resolution of the long-MRPC
along the long strip is about 0.6-1 cm. This satisfies the needs
for a large-area muon detector. The details of the long-MRPC
construction and its performance in the cosmic ray and beam tests
can be found in this paper~\cite{MTDNIMA}.

\section{Prototype performance of muon detector at STAR}\label{starresult}

\begin{figure}
\includegraphics[keepaspectratio,scale=0.5]{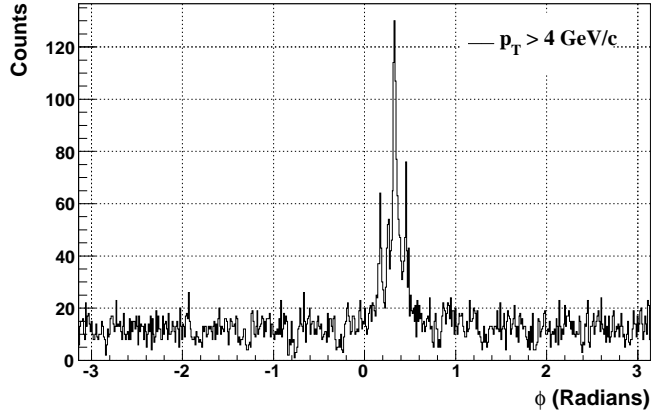}
\caption{(in color online) Azimuthal angle distribution of
particles of $p_{T}\!>\!4$ GeV/$c$ in Au+Au collisions,
extrapolated from the TPC to a radius of 400 cm.} \label{figure3}
\end{figure}

\begin{figure}
\includegraphics[keepaspectratio,scale=0.5]{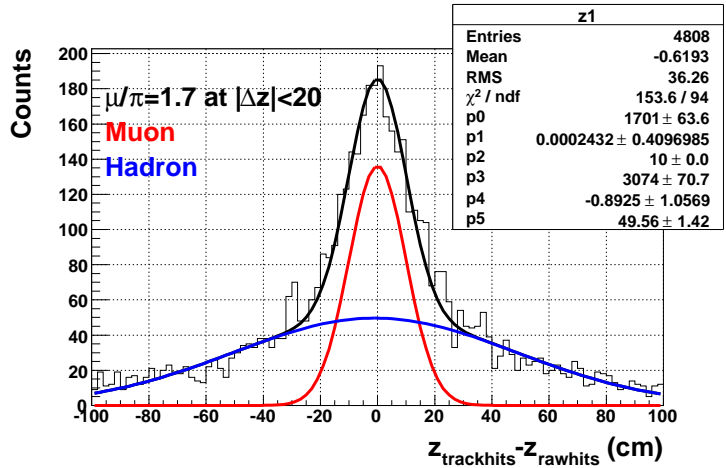}
\caption{(in color online) $\Delta$z distribution between
extrapolated hits and MTD hits in the MTD barrel.} \label{figure4}
\end{figure}

The prototype of the MTD, covered $\pi/60$ in azimuth and
$-0.25\!<\!\eta\!<\!0.25$ in pseudorapidity at a radius of
$\sim400$ cm during the 2007 run in 200 GeV Au+Au collisions. It
contained two long-MRPC modules. The prototype was placed outside
of the magnet steel that serves as hadron absorber. The prototype
successfully triggered the data acquisition system.
Fig.~\ref{figure3} shows azimuthal angle distribution of particles
from the TPC extrapolated to a radius of 400 cm in Au+Au
collisions at transverse momentum $p_{T}\!>\!4$ GeV/c. The peak
shows an enhancement of particle yield at the angle where MTD is
positioned. This indicates that offline tracking of particles from
the TPC was able to match hits from the Long-MRPC. The tracks of
the TPC were extrapolated to the MTD barrel, resulting in position
information from tracking. The time difference from two-end
readout of the hit strip gave us a position measurement along the
long strip of the long-MRPC. The difference of these two position
values in the z direction ($\Delta$z) is shown in
Fig.~\ref{figure4}, where the z direction is the beam direction.
Two components were observed in the $\Delta$z distribution. A
double Gaussian function was used to fit the distribution. The
$\sigma$ of the narrow Gaussian was found to be $\sim\!10$ cm by
selecting tracks of $p_{T}\!>\!2$ GeV/$c$ while the other Gaussian
is significantly broader. From the GEANT simulation, it shows that
muons of $p_{T}\sim\!2.5$ GeV/$c$ generated at the TPC center will
result in a Gaussian distribution with a sigma of 9 cm in the z
direction in the MTD barrel, after traversing the detector
material from the TPC center to the MTD. The simulation also
indicates that pions will result in a much broader distribution.
Assume the broad distribution is dominated by hadrons and narrow
Gaussian is dominated by muons, we obtained the muon to hadron
ratio is 1.7 and muon-to-hadron enhancement factor is about
200-300 at $\Delta z \!<\!20$ cm by requiring track matching only.
Additional dE/dx and time of flight cuts significantly enhanced
the muon-to-hadron ratio.

The average long-MRPC timing resolution for the two modules used
in this analysis was measured to be $\sim\!300$ ps in Au+Au
collisions. The ``start'' timing was provided by two identical
upgraded pseudo-vertex position detectors (upVPD), each 5.4 m away
from the TPC center along the beamline~\cite{upVPD}. After
subtracting the start timing jitter and detector material effect
contribution, the timing resolution from the MTD was found to be
not as good as those from cosmic and beam tests. This is
understood by the fact that the electronics we are currently using
are not designed for precise time measurement. With the proposed
full scale detector, we will improve our electronics.

\section{Conclusions}\label{concl}
In summary, research on a large-area, cost-effective muon
telescope detector has been carried out for STAR and for next
generation detectors at a future QCD Lab from state-of-the-art
multi-gap resistive plate chambers with large modules and long
strips. Cosmic ray and beam tests show the intrinsic timing
resolution of the long-MRPC is about 60-70 ps and spatial
resolution is better than 1 cm. The MTD triggered data at STAR
show that offline tracking of particles from the TPC was able to
match hits from the Long-MRPC. A clear muon peak was observed. The
hadron rejection power is found to be a few hundreds by requiring
track matching only.

In the year 2008 run, we took MTD triggered data in d+Au and p+p
collisions. Possible physics topics such as electron muon
correlations, muon spectra and elliptic flow will be pursued in
d+Au, p+p and Au+Au collisions. For the year 2009 run, we plan to
install another prototype tray, which will be equipped with the
same electronics as the time of flight system at STAR, to further
address the timing resolution of the Long-MRPC at STAR. We plan to
optimize the detector configuration and write a proposal for
full-coverage muon telescope detector at STAR.

\section*{Acknowledgments}
The author thank the Battelle Memorial Institute and Stony Brook
University for support in the form of the Gertrude and Maurice
Goldhaber Distinguished Fellowship.

\vfill\eject
\end{document}